\documentstyle[amssymb,preprint,12pt,aps,epsf]{revtex}
\begin{document}
\title{ESR study in lightly doped La$_{1-x}$Sr$_x$MnO$_3$}
\author{V. A. Ivanshin $^a$, J. Deisenhofer $^a$, H.-A. Krug von Nidda $^a$, A. Loidl
$^a$, A. A. Mukhin $^b$, A. M. Balbashov $^c$, and M. V. Eremin
$^d$}
\address{$^{a\text{ }}$Experimentalphysik V, EKM, Universit\"{a}t Augsburg, 86135\\
Augsburg, Germany\\ $^b$ Institute of General Physics, Russian
Academy of Sciences, 117942\\ Moscow, Russia\\ $^c$ Moscow Power
Engineering Institute, 105835 Moscow, Russia\\ $^d$ Kazan State
University, 420008 Kazan, Russia}
\date{\today}
\maketitle

\begin{abstract}
We present a systematic electron-spin-resonance (ESR) study in
single crystals of La$_{1-x}$Sr$_x$MnO$_3$ (0 $\leq x\leq$ 0.2).
The temperature dependence of the ESR linewidth marks all
significant transitions between both orthorhombic ($O\acute{}, O$)
and the rhombohedral ($R$) structural phases of the $T-x$ phase
diagram. All significant peculiarities of the ESR spectra for low
$x$ values ($x\leq 0.125)$ can be attributed to the cooperative
Jahn-Teller effect on the Mn$^{3+}$ e$_g$ states and to the
influence of the Dzyaloshinsky-Moriya exchange interaction.
Possible relaxation mechanisms at higher doping levels are
discussed.

\end{abstract}

\smallskip \newpage

\section{\protect\smallskip INTRODUCTION}

\smallskip
Due to the interplay of spin, charge, orbital, and lattice degrees
of freedom the doped perovskite manganites
(La$^{3+}_{1-x}$Me$^{2+}_x$Mn$^{3+}_{1-x}$Mn$^{4+}_x$O$_3$; Me =
Ca, Sr, Pb, and Ba \cite{jonker}) reveal a variety of interesting
phenomena, among which colossal magnetoresistance (CMR)
\cite{kusters,helmolt}, which also can be induced by an electric
field \cite{asamitsu}, magnetic-field induced structural phase
transitions \cite{tomioka} and photo-induced metal to insulator
transitions \cite{miyano} can be found. This variety of very
unusual phenomena corresponds to a complex (x, B, T)-phase diagram
exhibiting magnetic order, which strongly depends on the hole
concentration \cite{gennes,goodenough}, orbital order
\cite{goodenough,murakami}, charge order \cite{yamada} and
different structural phases, which depend on the tolerance factor
and on long-range Jahn-Teller distortions \cite{bogush}. The
ground-state properties of the manganites arise from the orbital
degeneracy with a strong Hund's coupling and are derived from a
strong competence of super-exchange \cite{goodenough} and
double-exchange \cite{zener} interactions and by the concomitant
appearance of charge order and long-range Jahn-Teller distortions.

A phase diagram for La$_{1-x}$Sr$_x$MnO$_3$ for strontium-doping
concentrations $x \leq$ 0.2 has been published recently by
Paraskevopoulos et al. \cite{para}. In accordance to an early
investigation \cite{bogush} the sequence rhombohedral $R$ to
orthorhombic $O$ (at a transition temperature of 1020 K) and then
to another orthorhombic $O\acute{}$ structure (at 760 K) was
observed in pure LaMnO$_3$, which is electrically insulating and
displays antiferromagnetic ordering of the Mn$^{3+}$ ions at the
Neel temperature $T_N$ = 140 K. The three octahedral Mn-O lengths
are strongly anisotropic in the $O\acute{}$ phase, due to a
long-range Jahn-Teller distortion, and become almost equal in the
$O$ phase. With increasing doping concentration $x$ the
conductivity of the manganites increases. This can be attributed
to the growing amount of holes in the lattice on Mn$^{4+}$ sites.
Owing to the strong Hund's coupling, the hopping of an e$_g$
electron between Mn$^{3+}$ and Mn$^{4+}$ sites is the dominating
mechanism for the conductivity of manganites in the paramagnetic
regime \cite{jaime}. Driven by an increasing concentration of
mobile holes, at low temperatures the insulating and
antiferromagnetic (AFM) structure passes to a purely metallic and
ferromagnetic state for $x > 0.17$. Close to $x = 1/8$ the ground
state is a ferromagnetic (FM) insulator. The Jahn-Teller
distortions of the $O\acute{}$ phase become suppressed for $x
> 0.15$, where the undistorted $O$ phase extends to the lowest
temperatures. The competition between ferromagnetism and
anitferromagnetism causes a complicated ground state. At the
moment, the coexistence of antiferromagnetic spin correlations
along with the conventional ferromagnetic spin correlations is
explained either within the framework of a canted
antiferromagnetic (CA) state or with a mixed two-phase state. The
existence of a phase separation was reported in the lightly doped
La$_{1-x}$Sr$_x$MnO$_3$ samples from a computational analysis in
the framework of the two-orbital Kondo model with Jahn-Teller
polarons \cite{yunoki}. A phase separation was found to occur
between (i) hole-undoped antiferromagnetic and hole-rich
paramagnetic regions and (ii) antiferomagnetic and ferromagnetic
phase with low, respectively high, hole concentration. Recent
neutron scattering\cite{yamada}, optical\cite{yung}, and NMR \cite
{papa} measurements on manganites can be interpreted in the
framework of this theoretical model.

The perovskite manganites show intense electron-spin-resonance
(ESR) signals with a large variation of the ESR-line parameters as
a function of the temperature \cite{causa}. Several controversial
approaches have been made to explain the ESR data in manganites.
In the paramagnetic state the ESR spectrum consists of a single
resonance line with a g value near 2.0. The broad ESR line in pure
LaMnO$_{3}$ (ESR linewidth $\Delta H \sim$ 2.5 kOe at room
temperature) was ascribed to Mn$^{3+}$ \cite{granado}. The linear
temperature dependence of the linewidth in
La$_{0.67}$Sr$_{0.33}$MnO$_3$ and
La$_{0.62}$Bi$_{0.05}$Ca$_{0.33}$MnO$_3$ was related to
contributions from spin-phonon relaxation \cite{seehra}. The
characteristic differences observed in ESR intensity and linewidth
in La$_{1-x}$Ca$_x$MnO$_3$ ($x$=0.1; 0.2) ceramic samples were
attributed to a model, in which a bottlenecked spin relaxation
takes place from Mn$^{4+}$ ions via Mn$^{3+}$ Jahn-Teller ions to
the lattice \cite{shengelaya}. The authors argue that the ESR
signal is due to Mn$^{4+}$ only, but renormalized by the exchange
coupling to Mn$^{3+}$. In ref. \cite{lofland} it was found that
all Mn spins contribute to the ESR signal in
La$_{1-x}$Sr$_x$MnO$_3$ ($x$ = 0.1; 0.2; 0.3) single crystals. The
authors suggest that the linewidth increases linearly as a result
of spin-lattice relaxation involving a single phonon. Finally,
Causa et al.\cite{causa} showed that all Mn spins contribute to
the resonance in perovskite compounds A$_{0.67}$Me
$_{0.33}$MnO$_3$ (A = La, Pr; Me = Ca, Sr) due to spin-spin
interactions, and not only the Mn$^{4+}$ or some spin clusters,
with no evidence of a spin-phonon contribution to the experimental
linewidth. Therefore, it is very important to perform ESR
measurements on manganites in an extended concentration and
temperature range in order to clarify the situation and to
correlate the results with different theoretical models. There are
extremely few publications devoted to the ESR studies on
manganites in the low doped region with only several doping
concentrations of $x$ $\leq $ 0.2
\cite{shengelaya,lofland,rettori,baginsk}.

In the present paper we report on our systematic ESR study in
La$_{1-x}$Sr$_x$MnO$_3$ ($x$ = 0; 0.05; 0.075; 0.1; 0.125; 0.15;
0.175; 0.2) single crystals in the paramagnetic regime in attempt
to compare the temperature dependence of the width and intensity
of the ESR lines with a detailed and complete phase diagram (see
Fig. \ref{phadia}), which was established recently in the same
samples after electrical conductivity, magnetic susceptibility,
submillimeter permittivity, and dynamic conductivity
investigations \cite{mukhin,para}. Particular attention is devoted
to the discussion of the linewidth in the Jahn-Teller distorted
phase.
\smallskip

\section{EXPERIMENTAL}

Single crystals of La$_{1-x}$Sr$_x$MnO$_3$ were grown by the
floating-zone method with radiation heating in air atmosphere, as
described in ref. \cite{mukhin}. The purity of the chemicals used
for sample preparation (La$_2$O$_3$, SrCO$_3$ and Mn$_3$O$_4$) was
not less than 99.99\%. X-ray powder diffraction proved that the
grown materials were of single phase. However, two dimensional
X-ray TV topography of the crystals has revealed a twin structure.

ESR measurements were performed with a Bruker ELEXSYS E500
CW-spectrometer at X-band frequencies ($\nu \approx$ 9.2 GHz)
equipped with continuous gas-flow cryostats for He (Oxford
Instruments) and N$_2$ (Bruker) in the temperature range between
4.2 K and 680 K. Small crystals of 0.6 - 3.7 mg were chosen for
our ESR experiments to avoid the sample-size effects \cite{causa}.
The samples were placed into quartz tubes and fixed with either
paraffin (at low temperatures 4 K $\leq T \leq$ 300 K) or NaCl (at
300 K $\leq T \leq$ 680 K). We have restricted our experiments to
the paramagnetic regime ($T > T_N$) only, because within the
magnetically ordered regime the ESR behavior is very sample
dependent, which is attributed to magnetic inhomogeneity of the
samples in the antiferromagnetic and ferromagnetic phases due to
local variations of oxygen stoichiometry and chemical composition
\cite{causa,lofland}.

\subsection{ESR\ spectra}

Electron-spin resonance detects the power $P$ absorbed by the
sample from the transverse magnetic microwave field as a function
of the static magnetic field $H$. The signal-to-noise ratio of the
spectra is improved by recording the derivative $dP/dH$ with
lock-in technique. ESR spectra, which are characteristic for the
three structural phases of the paramagnetic regime, are presented
in Fig. \ref{spectra}, illustrating their evolution with Sr
concentration $x$ (left column) and temperature $T$ (right
column). Within the whole paramagnetic regime the spectrum
consists of a broad, exchange narrowed resonance line, which is
well fitted by a Dysonian line shape \cite{feher}. As the
linewidth $\Delta H$ is of the same order of magnitude as the
resonance field $H_{res}$ in the present compounds, one has to
take into account both circular components of the exciting
linearly polarized microwave field. Therefore the resonance at the
reversed magnetic field $-H_{res}$ has to be included into the fit
formula for the ESR signal, given by
\begin{equation}
\frac{dP}{dH} \propto \frac{d}{dH}(\frac{\Delta H + \alpha
(H-H_{res})}{(H-H_{res})^2 + \Delta H^2} + \frac{\Delta H + \alpha
(H+H_{res})}{(H+H_{res})^2 + \Delta H^2})
\label{dyson}
\end{equation}
This is an asymmetric Lorentzian line, which includes both
absorption and dispersion, where $\alpha$ denotes the
dispersion-to-absorption ratio. Such asymmetric line shapes are
usually observed in metals, where the skin effect drives electric
and magnetic microwave components out of phase in the sample and
therefore leads to an admixture of dispersion into the absorption
spectra. For samples small compared to the skin depth one expects
a symmetric absorption spectrum ($\alpha$ = 0), whereas for
samples large compared to the skin depth absorption and dispersion
are of equal strength yielding an asymmetric resonance line
($\alpha$ = 1).

At low Sr concentrations or at low temperatures the spectra are
nearly symmetric with respect to the resonance field in accordance
with pure absorption spectra with $\alpha$=0. With increasing $x$
or $T$ they become more and more asymmetric corresponding to an
increasing parameter $\alpha$. To check, whether the skin effect
is the reason for the asymmetric line shape, we have to estimate
the skin depth $\delta = (\rho / \mu_0 \omega)^{0.5}$ from the
electric resistance $\rho$ and the microwave frequency $\omega = 2
\pi \times 9$ GHz ($\mu_0 = 4 \pi \times 10^{-7}$ Vs/Am). Using
the resistance values determined by Mukhin \cite{mukhin} - for
example $\rho$ = 0.2 $\Omega$ cm for $x = 0.125$ at room
temperature - we find a skin depth $\delta =$ 0.16 mm. Within the
whole paramagnetic regime, the resistance values decrease by at
least two orders of magnitude with increasing temperature or
increasing Sr concentration. Therefore the skin depth is larger
than the sample dimensions at low $T$ or $x$ and becomes
comparable to or even smaller than the thickness of the sample as
these parameters increase in accordance with the increasing
asymmetry of the ESR spectra.

The resonance field of all compounds under investigation yields a
g value slightly below the free-electron value, which is usual for
transition-metal ions with a less than half filled d shell
\cite{abragham}. This g value is found to be nearly isotropic $g
\approx 1.98$ in both the orthorhombic $O$ and the rhombohedral
$R$ phase, but it reveals a weak anisotropy 1.94 $\leq g \leq$
1.98 within the Jahn-Teller distorted $O\acute{}$ phase. Finally
approaching the ordering temperature from above, the whole
resonance becomes seriously distorted and is strongly shifted to
lower or higher fields dependent on the orientation of the sample.
This is caused by internal fields due to the onset of magnetic
order. More detailed information on the different phases is
obtained from the resonance linewidth and the ESR intensity, which
are presented in the following subsection.

\subsection{Temperature dependence of the ESR linewidth and intensity}
In Fig. \ref{HWHMvsT} the ESR linewidth $\Delta H$ is plotted
versus temperature for all investigated samples. Here the
orientation of the single crystals was adjusted to the most
pronounced temperature dependence, which could be found within the
Jahn-Teller distorted $O\acute{}$ phase and is described in more
detail below. Only for the pure sample $x = 0$ it was impossible
to find an unequivocal orientation as it is presumably a twinned
crystal. On the other hand, the crystals with Sr concentrations $x
\ge 0.15$ had not been oriented, because they never reach the
Jahn-Teller distorted regime and their paramagnetic spectra are
isotropic.

The magnitude of $\Delta H$ is approximately constant in the whole
temperature range 150 K $< T <$ 650 K in pure LaMnO$_3$, because
the structural transition into the $O$ phase takes place close to
800 K \cite{mukhin}. However, in all doped samples the temperature
dependence of $\Delta H$ exhibits a number of deviations from a
monotonic behavior, which we identified with different magnetic
and structural phase transitions. At lowest temperatures of
observation (150 K $< T <$ 200 K) $\Delta H$ passes a minimum, as
the temperature approaches $T_N$ from above, and then increases
abruptly near $T_N$. This dependence indicates the transition from
the paramagnetic regime into the magnetically ordered structure
and is connected with the strong shift of the resonance field
mentioned above. The typical features of the linewidth obtained
within the different paramagnetic regimes can be summarized as
follows:

(a) Orthorhombic strongly Jahn-Teller distorted $O\acute{}$ phase:
A strongly anisotropic linewidth was observed in the lightly doped
samples with $x \leq 0.125$. Fig. \ref{dHsr005} depicts the
temperature dependence of the linewidth for the three main
orientations of the external magnetic field $H$ in
La$_{0.95}$Sr$_{0.05}$MnO$_3$. If the magnetic field is found
within the a-b plane, the linewidth passes a maximum with
increasing temperature, whereas the linewidth remains nearly
constant, if the field is parallel to the c axis. At $T$ = 600 K,
where the structural transition to the $O$ phase takes place, the
values of $\Delta H$ coincide for all directions. The angular
dependence of the linewidth, which is shown in Fig. \ref{angle},
can be commonly described by $A+B\sin^2\vartheta$. At the
temperature of maximal anisotropy ($T$ = 400 K) we found the ratio
$B/A = 0.56$.

(b) Orthorhombic $O$ phase: The upper temperature limit of the
anisotropy in the linewidth is changing in the $Sr$ doped samples
from 600 K ($x$ = 0.05) to 260 K ($x$ = 0.125). This coincides
quite well with the $O\acute{}\longrightarrow O$ transition from
the strongly Jahn-Teller distorted orthorhombic structure to the
weakly distorted orthorhombic pseudocubic one. In this phase, the
linewidth $\Delta H$ exhibits a linear temperature dependence with
a typical slope of about 2 Oe/K.

(c) Rhombohedral $R$ phase: Again, a linear temperature dependence
of the linewidth is observed for all samples investigated in this
phase ($0.075 \leq x \leq 0.2$) up to 600 K. Its slope is
comparable to the $O$ phase. For $x < 1.5$ there is a slight kink
at the $O\longrightarrow R$ transition. Finally, the data points
at the highest temperatures of this investigation $(600 K < T <
680 K)$ indicate a tendency to saturation of $\Delta H$.

The integrated intensity $I(T)$ of the resonance line measures the
spin susceptibility $\chi_{ESR}$ of the ESR probe. For
ferromagnetically coupled ions its temperature dependence usually
follows a ($T-\Theta_{cw})^{-1}$ Curie-Weiss law, where $\Theta
_{CW}$ is the Curie-Weiss temperature. If all spins take part in
producing the ESR signal, it is reasonable to plot $1/I$ versus
$T$ and to expect a linear behavior all over the paramagnetic
regime. However, as one can obtain from Fig. \ref{intens}, this
procedure does not yield a unique linear temperature dependence at
all. Only LaMnO$_3$ behaves linearely with a Curie-Weiss
temperature $\Theta_{CW}(x = 0) \approx$ 87 K in the complete
temperature range under consideration. For Sr concentrations 0.05
$\leq x \leq$ 0.125 we observe a distinct kink near the structural
phase transition $O\acute{}\longrightarrow O$, just at the
temperature where the anisotropy of the linewidth starts to
decrease with increasing temperature. Below this kink, the inverse
intensity shows a linear temperature dependence with a similar
slope as in LaMnO$_3$. The respective Curie-Weiss temperatures
increase with increasing Sr concentration up to $\Theta_{CW}(x =
0.125) \approx$ 208 K. To analyze the origin of the ESR signal, we
compared its intensity at temperatures below the kink to the ESR
intensity of the so called green phase Gd$_2$BaCuO$_5$, which
exhibits an ESR signal with a similar linewidth. The respective
compound, in which all Gd$^{3+}$ spins contribute to the ESR
signal, shows a Curie-Weiss susceptibility with $\Theta_{CW} =
-23$ K \cite{goya}. Starting with pure LaMnO$_3$, we found that
all Mn$^{3+}$ ions contribute to the ESR signal with their full
magnetic moment in the $O\acute{}$ phase. With increasing Sr
concentration $x$ the averaged magnetic moment of both Mn$^{3+}$
and Mn$^{4+}$ ions even increases slightly by nearly 10 percent up
to $x = 0.125$ within this phase, although Mn$^{4+}$ ions (spin S
= 3/2) exhibit a smaller magnetic moment than Mn$^{3+}$ ions (spin
S = 2). Above the kink the inverse intensity again reaches a
linear temperature dependence with a remarkably higher slope than
below the kink (about 13 and 8 times the low-temperature slope for
x = 0.05 and 0.1 respectively). As the inverse slope measures the
total number of spins, which generate the ESR signal, about 90
percent of the spins seem to become invisible at the transition to
the $O$ phase, suggesting that the contribution of all Mn$^{3+}$
ions disappears. However, we have to take the skin effect into
account once more. Comparing the temperature of the kink in the
intensity with the resistance data \cite{mukhin}, we realize that
it approximately coincides with the temperature, where the
specific resistance drops below 1 $\Omega$ cm with increasing
temperature. This corresponds to a skin depth of about 0.4 mm,
which is of the order of the sample size. Therefore, the part of
the crystal, which is accessible for the microwave field and
contributes to the ESR signal, is more and more reduced with
increasing temperature. To check the importance of the skin effect
in our experiments, we investigated a powder sample with a Sr
concentration $x = 0.1$. Its inverse ESR intensity still showed a
discontinuity at the phase transition $O\acute{}\longrightarrow O$
but at higher temperatures it attained nearly the same slope as at
low temperatures, indicating that all Mn spins contribute to the
ESR signal in the $O$ phase, as well.

\section{DISCUSSION}
\subsection{Intensity of ESR}
Before we analyze the linewidth data in detail, we first focus on
the ESR intensity in order to show that its temperature dependence
is in accordance with the behavior expected for Mn$^{3+}$ within
an intermediate octahedral ligand field under the influence of
distortion \cite{abragham}: The octahedral field splits the five d
orbitals into a lower triplet t$_{2g}$ and an upper doublet e$_g$
with an energy splitting of about 10$^4$ K. In the case of
intermediate ligand fields the dominating Hund's coupling causes a
parallel spin alignment of the four d electrons, where three of
them occupy the triplet t$_{2g}$ and the forth is found in the
doublet e$_g$ (so called high-spin state S=2). Therefore the
ground state is an orbital doublet $\Gamma_3$. Including the spin,
the $\Gamma_3$ state is tenfold degenerated. This degeneration is
lifted by second order spin-orbit couplings into a manifold of
singlet, doublet, and triplet states with an overall splitting of
about 20 K \cite{abragham}. For this case the Zeeman effect is
very complicated and the observation of an ESR signal is rather
unlikely. The situation changes drastically for a tetragonal
Jahn-Teller distortion, which splits the orbital doublet
$\Gamma_3$ into two singlets with a typical energy gap of the
order 10$^3$ K. Then the ground state is a spin quintuplet with
usual Zeeman effect. Taking the spin-orbit coupling into account,
the ESR signal is expected at about $g_z \approx 1.95$ and $g_x =
g_y \approx 1.99$ \cite{abragham}, which agrees very well with our
experimental data. In the case of Mn$^{4+}$ in an octahedral
field, the three d electrons occupy the triplet t$_{2g}$ with
parallel spin alignment S = 3/2. The ground state is an orbital
singlet $\Gamma_2$ with a spin quadruplet. This yields an ESR
signal with an isotropic g value at about $g \approx 1.99$.

Therefore the ESR signal is due to all Mn$^{3+}$ and Mn$^{4+}$
spins within the Jahn-Teller distorted $O\acute{}$ phase. The
increase of the averaged moment with the Sr concentration $x$,
which is found in contrast to the expected linear superposition of
Mn$^{3+}$ and Mn$^{4+}$ spins, agrees with susceptibility
measurements \cite{para} and may be ascribed to ferromagnetic
polarization around the Mn$^{4+}$ spins. With suppression of the
Jahn-Teller distortion at the transition $O\acute{}\longrightarrow
O$ the contribution of the Mn$^{3+}$ ions should vanish, if their
surrounding is ideally octahedral. Obviously the weak orthorhombic
distortion of the $O$ phase is still strong enough to retain the
ESR signal of Mn$^{3+}$.

\subsection{Linewidth}
The result that the Jahn-Teller Mn$^{3+}$ ions dominate the ESR
signal within the $O\acute{}$ phase is essential for the following
discussion of the linewidth within this regime: All significant
peculiarities of the ESR spectra in the respective phase, i. e.
the magnitude of the ESR linewidth, its anisotropy, and the
anisotropy of the resonance field, are very similar to those
observed by Tanaka et al. \cite{tanaka} in the quasi-one
dimensional compound CsCuCl$_3$. That system exhibits a
cooperative Jahn-Teller effect in the temperature range between
120 K and 560 K. The authors explain their ESR results due to
Dzyaloshinsky-Moriya (DM) antisymmetric interaction \cite{moriya}
between Jahn-Teller Cu$^{2+}$ ions i and j, given by
\begin{equation}
{\cal H}_{ij} = \bold{D_{ij}}\cdot[\bold{S_i} \times  \bold{S_j}].
\label{DMham}
\end{equation}
ESR investigations in KCuF$_3$ \cite{yama} and CuGeO$_3$
\cite{yamad}, which are quasi-one dimensional systems as well,
have been treated in a similar way. In those publications it was
shown that the usual dipolar or anisotropic exchange interactions
cannot produce the observed linewidth of the order of 10$^3$ Oe,
because the strong exchange interaction between the Cu$^{2+}$ ions
leads to a considerable narrowing of the ESR line. The present
compound La$_{1-x}$Sr$_x$MnO$_3$ exhibits comparable exchange
interactions between Mn$^{3+}$ ions (for $x$ = 0) with spin $S$ =
2 and between Mn$^{3+}$ and Mn$^{4+}$ ions with $S$ = 3/2 (for $x
> 0$). Therefore, we suggest that the broad ESR line in the
strongly Jahn-Teller distorted $O\acute{}$ phase of LaMnO$_3$ can
be interpreted in terms of the DM antisymmetric interaction
\cite{moriya} between Mn$^{3+}$ ions, which are strongly
super-exchange coupled via oxygen ions. The dominancy of the
antisymmetric interactions in the Jahn-Teller distorted phase is
also confirmed by the fact that non-Jahn-Teller Mn$^{4+}$ ions
suppress the line-broadening for $x > 0.125$. This finding
supports the suggestion of Millis \cite{millis} that strong
coupling between charges and lattice is observed in the Sr doped
manganites due to the cooperative Jahn-Teller effect for Sr
concentrations less than $x \approx$ 0.15.

Looking at the magnetic structure \cite{huang} of the
antiferromagnetic phase (see Fig. \ref{struc}), one recognizes
that each Mn$^{3+}$ ion is coupled ferromagnetically to its four
next Mn$^{3+}$ neighbors within the a-b plane and
antiferromagnetically to the other two in direction of the c axis.
Therefore the compound is quasi-one dimensional along the
antiferromagnetically coupled Mn$^{3+}$ chains in c direction, and
the analysis can be performed similar to the compounds cited
above. Following Yamada \cite{yama}, for temperatures $k_BT$ large
compared to the exchange interaction $J$ between the Mn ions, the
linewidth contribution $\Delta H_{DM}$ of the DM interaction is
approximated by
\begin{equation}
\Delta H_{DM} = \frac{{\bold{D_{ij}}}^2 S(S+1)}{6 g \mu _B J}
f(\vartheta), \label{dHDM}
\end{equation}
where $f(\vartheta)$ = (1 + $\cos^2\vartheta$)/4 for
$\bold{D_{ij}}$ parallel or $f(\vartheta)$ = (2 +
$\sin^2\theta$)/8 for $\bold{D_{ij}}$ perpendicular to the
quasi-one dimensional chain direction. Furthermore the g-shift
$\Delta g = g - 2$ is given by
\begin{equation}
\Delta g = \frac{D}{2J} g, \label{dgDM}
\end{equation}
where $D$ and $g$ denote the absolute value of $\bold{D_{ij}}$ and
the mean value of the anisotropic g tensor, respectively.

From Fig. \ref{dHsr005} and Fig. \ref{angle} we obtain that for
high temperatures $T >$ 300 K the linewidth changes only slightly
within the a-b plane but is strongly anisotropic with respect to
the c axis following a $\sin^2\vartheta$ law. As it is shown in
the insert of Fig. \ref{dHsr005}, the anisotropy reaches a maximum
value of $\Delta H_a/\Delta H_c = 1.56$. This agrees quite well
with the expected value of 1.5 for $f(\vartheta)$ = (2 +
$\sin^2\vartheta$)/8. So the vector $\bold{D_{ij}}$ is pointing
perpendicular to the quasi-one dimensional chain direction. To
identify the origin of the DM interaction in the present system
and to estimate its magnitude, we have to take into account the
distortions of the crystal structure (see Fig. \ref{struc}).
Indeed, according to microscopic theory \cite{keffer,moskvin}, one
obtains
\begin{equation}
\bold{D_{ij}} = D_O\cdot[\bold{n_{iO}} \times  \bold{n_{Oj}}],
\label{DMv}
\end{equation}
where $\bold{n}$ are unit vectors along Mn$^{3+}$-O$^{2-}$ bonds.
In LaMnO$_3$ these bonds are found along the chain direction
\cite{huang}, as indicated in Fig. \ref{struc}, and the bridge
angle $\phi$ is about 155$^{\circ}$. Therefore the vector product
[$\bold{n_{iO}} \times \bold{n_{Oj}}$] does not vanish and
$\bold{D_{ij}}$ is perpendicular to the
Mn$^{3+}$-O$^{2-}$-Mn$^{3+}$ triangle plane and to the chain axis.
Taking into account the Jahn-Teller effect at the Mn$^{3+}$ site,
we conclude that the super-exchange coupling is mainly realized
via $3z^2-r^2$ orbitals of the Mn$^{3+}$ ions. The situation is
very similar to that which occurs in RFeO$_3$, where R denotes a
rare-earth ion, because the orbitals $x^2-y^2$ are not active in
the super-exchange coupling of Fe$^{3+}$-O$^{2-}$-Fe$^{3+}$, too.
Using the channel model \cite{eremin}, we get
\begin{equation}
J_{Mn^{3+}-O^{2-}-Mn^{3+}} = \frac{S^2_{Fe^{3+}}}{S^2_{Mn^{3+}}}
J_{Fe^{3+}-O^{2-}-Fe^{3+}} \label{Jmn}
\end{equation}
Furthermore we can estimate the parameter of the DM interaction in
the same manner \cite{eremi}
\begin{equation}
D_{Mn^{3+}-O^{2-}-Mn^{3+}} = \frac{S^2_{Fe^{3+}}}{S^2_{Mn^{3+}}}
D_{Fe^{3+}-O^{2-}-Fe^{3+}} \label{Dmn}
\end{equation}
Using the data for Fe$^{3+}$ from ref. \cite{moskvin} and taking
into account the dependencies of $J$ and $D$ on the bridge angle
$\phi$ we finally get for Mn$^{3+}$ $J$ = 67 K and $D$ = 2.54 K.
With these values we estimate the expected g-shift $\Delta g =
0.04$ and linewidth $\Delta H = 2.15$ kOe, which agree quite well
with the experimental results.

Finally we shortly comment on the temperature dependence of the
linewidth in both $O$ and $R$ phases, which has been discussed
controversially by several authors as mentioned in the
introduction. For intermediate Sr concentrations 0.075 $\leq $ $x$
$\leq $ 0.15 the linear increase of the linewidth may be due to
the spin-lattice relaxation, as it was suggested in ref.
\cite{lofland}, because of the importance of electron-phonon
interactions at this doping level \cite{uhlen}. Nevertheless, the
quasi-linear behaviour of $\Delta H$ at $x > 0.15$ is also well
described by a model on the basis of effective Heisenberg-like
interactions for Mn$^{3+}$- Mn$^{4+}$ spin pairs with no evidence
of spin-phonon contribution to the experimental linewidth
\cite{causa}. The infinite-temperature linewidth for different
doped manganites with Me-concentration $x$ = 0.33 observed in the
latter study up to 1000 K was kept as adjustable parameter, which
is related to spin-only interactions.

Very recent studies of Shengelaya et al.\cite{shen} notice the
similarity between temperature dependence of ESR linewidth and
conductivity in manganites, which is determined by the hopping
motion of small polarons \cite{jaime}. Quasioptical spectroscopy
measurements on the same samples as in our studies reveal that
hopping or tunneling between localized states dominates the
conductivity for Sr concentrations $x \leq$ 0.15 and temperatures
$T <$ 300 K \cite{pimenov}. Even for
La$_{0.825}$Sr$_{0.175}$MnO$_3$ the localization effects are
observed at least 10 K below the magnetic phase transition.
Therefore, one can support that the hopping rate of e$_g$ charge
carriers defines a broadening dynamics of the ESR line.

\section{CONCLUSION}
We performed systematic ESR investigations on
La$_{1-x}$Sr$_x$MnO$_3$ single crystals for Sr concentrations (0
$\leq x \leq$ 0.2) within the paramagnetic regime. The ESR
properties of the strongly Jahn-Teller distorted $O\acute{}$ phase
have been investigated in detail for the first time. Comparing the
ESR intensity with Gd$_2$BaCuO$_5$, we obtained that the strongly
exchange narrowed paramagnetic resonance is due to both Mn$^{3+}$
and Mn$^{4+}$ ions within both the strongly Jahn-Teller distorted
$O\acute{}$ phase and the weakly distorted pseudocubic $O$ phase.
The temperature dependence of the ESR linewidth marks all
significant transitions between the structural phases of the $T-x$
phase diagram. Within the Jahn-Teller distorted $O\acute{}$ phase,
its magnitude and anisotropy are well understood in terms of the
exchange modulation of the Dzyaloshinsky-Moriya antisymmetric
interaction. The linewidth directly probes the suppression of the
Jahn-Teller distortion with increasing Sr concentration and
increasing temperature. In principle it should be possible to
obtain the distortion angle from these data. However, further
theoretical effort is necessary to describe the detailed
temperature and concentration dependence of the ESR parameters
under Jahn-Teller distortion.

\section*{ACKNOWLEDGEMENTS}
We thank B. Elschner, B. I. Kochelaev, V. K. Voronkova, A. E.
Usachev, A. Shengelaya and K. Held for useful discussions. This
work was supported by the Bundesministerium f\"{u}r Bildung und
Forschung (BMBF) under Contract No. 13N6917/0. M. V. Eremin was
supported in part by the Russian Foundation for Basic Research,
Grant No. 97-02-18598.

\begin{figure}
\caption{Phase diagram of La$_{1-x}$Sr$_x$MnO$_3$ from ref.
\protect\cite{para}: Structural phases $R$, $O$, $O\acute{}$,
$O''$; PM = paramagnetic, FM = ferromagnetic, CA = canted
antiferromagnetic; M = metallic, I = insulating.}
\label{phadia}
\end{figure}

\begin{figure}
\caption{ESR spectra of La$_{1-x}$Sr$_x$MnO$_3$. Left column:
various Sr concentrations $x$ at $T$ = 340 K ($x$ =0.2 in the
rhombohedral $R$ phase, top panel; $x$ =0.125 in the orthorhombic
$O$ phase, middle panel; $x$ =0.075 in the strongly distorted
orthorhombic $O\acute{}$ phase, down panel). Right column:
temperature evolution of the ESR spectrum for $x = 0.1$ in three
different structural phases. Solid lines represent the fits using
the Dysonian line shape, equation \protect\ref{dyson}.}
\label{spectra}
\end{figure}

\begin{figure}
\caption{Temperature dependence of the ESR linewidth $\Delta H$ in
La$_{1-x}$Sr$_x$MnO$_3$ (0 $\leq x \leq$ 0.2).}
\label{HWHMvsT}
\end{figure}

\begin{figure}
\caption{Temperature dependence of the ESR linewidth in
La$_{0.95}$Sr$_{0.05}$MnO$_3$ for the three main orientations of
$H$. The insert displays the relative anisotropy.}
\label{dHsr005}
\end{figure}

\begin{figure}
\caption{Angular dependence of the linewidth $\Delta H$ in
La$_{0.95}$Sr$_{0.05}$MnO$_3$ measured at 200 K with external
magnetic field $H \perp a$. The solid line represents a fit
$A+B\sin^2\vartheta$ as described in the text. The insert sketches
the Jahn-Teller distortion (cf. Fig. \protect\ref{struc}).}
\label{angle}
\end{figure}

\begin{figure}
\caption{Reciprocal intensity $1/\chi_{ESR}$ versus temperature in
La$_{1-x}$Sr$_{x}$MnO$_3$ for representative Sr concentrations
$x$.}
\label{intens}
\end{figure}

\begin{figure}
\caption{Crystallographic structure of LaMnO$_3$ in the strongly
Jahn-Teller distorted $O\acute{}$ phase \protect\cite{huang}. The
arrows on the Mn$^{3+}$ ions represent the antiferromagnetic
structure below 140 K.}
\label{struc}
\end{figure}

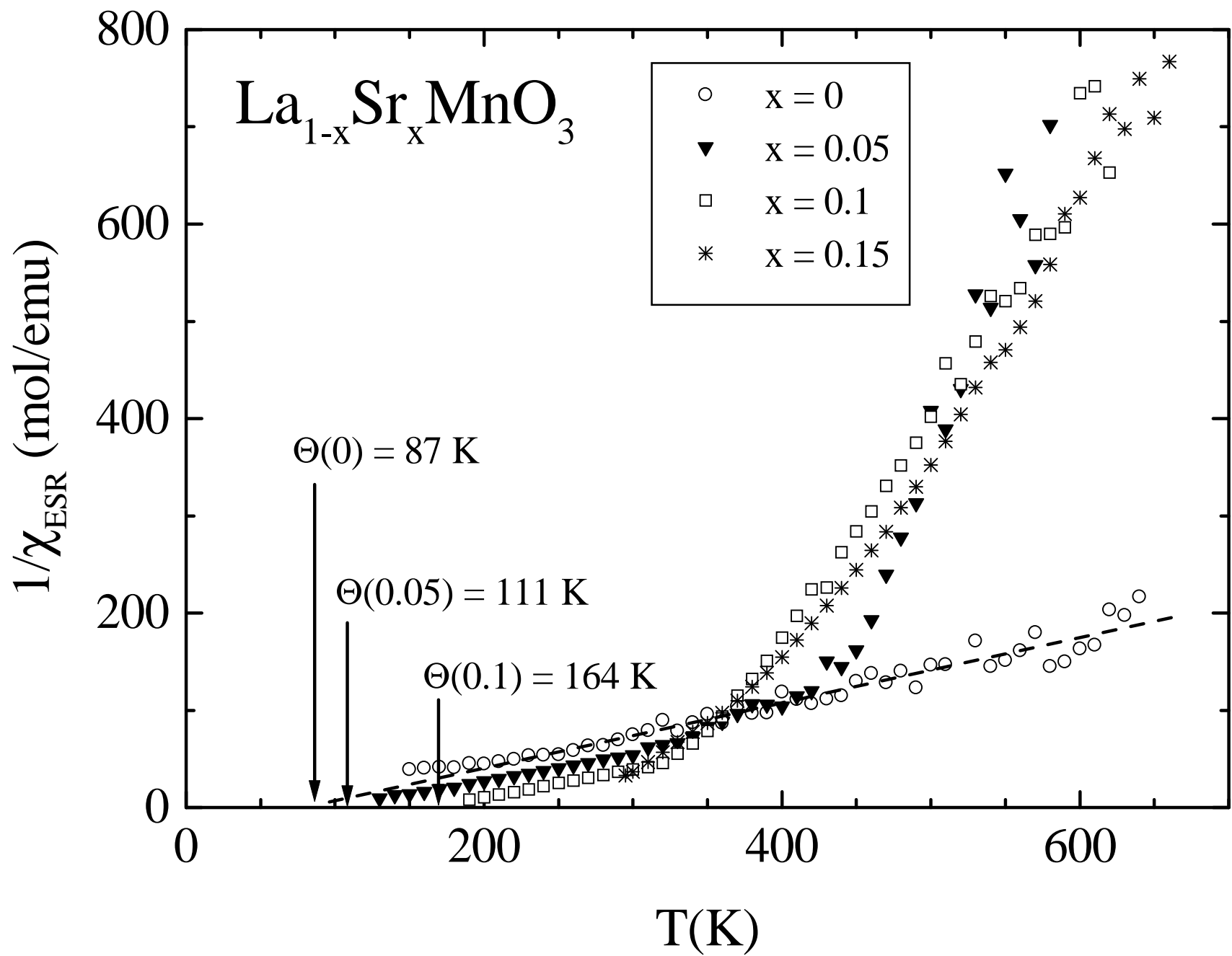
\end{document}